\documentclass[showpacs,twocolumn,aps,preprintnumbers,letterpaper]{revtex4}
\usepackage{amsmath,amssymb}
\usepackage{epsfig}
\usepackage{graphicx}
\usepackage{amsmath}
\usepackage{slashed}
\usepackage{amsfonts}
\usepackage{epstopdf}

\newcommand{\be}{\begin{equation}}
\newcommand{\ee}{\end{equation}}
\newcommand{\bear}{\begin{eqnarray}}
\newcommand{\eear}{\end{eqnarray}}
\newcommand{\ba}{\begin{array}}
\newcommand{\ea}{\end{array}}

\def\be{\begin{eqnarray}}
\def\ee{\end{eqnarray}}
\def\bea{\be}
\def\eea{\ee}

\def\roughly#1{\mathrel{\raise.3ex\hbox{$#1$\kern-.75em%
\lower1ex\hbox{$\sim$}}}}

\begin{document}

\title{Hydrodynamical Description of the\\
QCD Dirac Spectrum at Finite Chemical Potential}

\author{Yizhuang Liu}
\email{yizhuang.liu@stonybrook.edu}
\affiliation{Department of Physics and Astronomy, Stony Brook University, Stony Brook, New York 11794-3800, USA}

\author{Piotr Warcho\l{}}
\email{piotr.warchol@uj.edu.pl}
\affiliation{M. Smoluchowski Institute of Physics, Jagiellonian University, PL-30348 Krakow, Poland}

\author{Ismail Zahed}
\email{ismail.zahed@stonybrook.edu}
\affiliation{{}Department of Physics and Astronomy, Stony Brook University, Stony Brook, New York 11794-3800, USA}


\date{\today}
\begin{abstract}
We present a hydrodynamical description of the QCD Dirac spectrum at finite chemical
potential  as an uncompressible  droplet in the complex eigenvalue space.  For a large droplet, the fluctuation spectrum around the
hydrostatic solution is gapped by a longitudinal Coulomb plasmon, and exhibits a frictionless odd 
viscosity.  The stochastic relaxation  time for the restoration/breaking of chiral symmetry is set by twice the plasmon frequency.
The leading droplet size correction to the relaxation time is fixed  by a universal odd viscosity to density ratio
$\eta_O/\rho_0=(\beta-2)/4$ for the three Dyson ensembles $\beta=1,2,4$.
\end{abstract}






\maketitle

\setcounter{footnote}{0}



{\bf 1. Introduction.\,\,} 
QCD breaks spontaneously chiral symmetry with the emergence of an octet of light mesons
that permeate most of the hadronic processes at low energies~\cite{BOOK}. Dedicated lattice
simulations  are now in full support of this spontaneous 
breaking~\cite{LATTICE}.  Fundamental light quarks become constitutive and heavy
producing most of the mass of the elements around us.

A remarkable feature of the spontaneous breaking of chiral symmetry is the large accumulation of the eigenvalues
of the Dirac operator near zero-virtuality with the formation of a finite vacuum chiral condensate~\cite{CASHER}. 
Small eigenvalue virtuality translates to large proper time, as light quarks travel
very long in proper time and delocalize. The zero virtuality regime is ergodic, and its neighborhood is diffusive~\cite{DISORDER}.
This behavior is analogous to disordered electrons in mesoscopic systems~\cite{MONT}.

The ergodic regime of the QCD Dirac spectrum with its universal
spectral oscillation is described by a chiral random matrix model~\cite{SHURYAK}.  In short, the model simplifies
the Dirac spectrum to its zero-mode-zone (ZMZ). The Dirac matrix is composed of hopping between 
N-zero modes and N-anti-zero modes because of chirality, which are sampled from gaussian
ensembles thanks to the central limit theorem. The model was initially  suggested as a null dynamical
limit  of  the instanton liquid model~\cite{MACRO}.

QCD at finite chemical potential $\mu$ is notoriously difficult to sample on a lattice due to the sign problem
\cite{SIGN}.  A number of chiral models have been proposed to describe the effects of matter in QCD with
light quarks~\cite{BOOK}. In vacuum, the chiral random matrix model simplifies the QCD Dirac spectrum
to its ZMZ. In matter, the light quark zero modes are involved. Their chiral and cross-hopping in the ZMZ 
is suppressed exponentially, and the corresponding  Dirac matrix  is banded and not random. However,
large matter effects reduce the banded matrix to its diagonal, localizing the quark zero modes
into molecules. In the 1-matrix model the chiral random ensemble is deformed by a constant matrix, 
leading to localization at large $\mu$~\cite{STEPHANOV,US}. In the 2-matrix 
model the deformation is still  random and only generic for moderate $\mu$ with no strict
banding at large $\mu$~\cite{OSB,AKE}. 
The 1-matrix approach to QCD at finite $\mu$ has been discussed by many~\cite{BOOK,BLUE,RMANY}.

In this letter we would like to combine the ergodic  character of the chiral random matrix model 
for the low-lying modes in the ZMZ with the universal character of the hydrodynamics approach,
to describe the relaxation of the QCD Dirac eigenvalues at finite $\mu$.
We will obtain the following new results:
1/ a hydrodynamical description of the Dirac eigenvalues as a droplet in the complex 2-plane;
2/ small amplitude deformations in the droplet that are gapped by the emergence of a plasmon
 with an odd viscosity;
3/ a non-perturbative  estimate of the stochastic relaxation time for breaking/restoring chiral  symmetry in matter.

\vskip 0.5cm

{\bf 2. The model.\,\,}
A  useful model for 
the eigenvalues of the QCD Dirac operator at finite $\mu$ makes use of a 2-matrix model~\cite{OSB,AKE}

\bea
Z_\beta[m_f]=&&
\int \prod_{i=1}^Nd^2z_i|z_i|^{\alpha}\prod _{i<j}^N|z_i^2-z_j^2|^\beta\,\nonumber\\
&&\times(z_i^2+m_f^2)^{N_f}\,e^{-W(z_i)}
\label{1}
\eea
for quarks in the complex representation or $\beta=2$ with $\alpha=\beta(\nu+1)-1$. 
$\nu$ accounts for the difference between 
the number of zero modes and anti-zero modes. The potential is

\be
W(z)= \frac {Na\beta}{2l^2}\left(|z|^2-\frac{\tau}2 (z^2+{\overline z}^2\right)
\label{2}
\ee
with $l^2\equiv {1-\tau}=2\mu^2/(1+\mu^2)$. 
 For $\mu\rightarrow 0$, 
$\tau\approx 1$ and $l^2\approx 2\mu^2$, so that 
$W(z)\approx -(N/\mu^2)(z-\overline z)^2$,
which restricts the eigenvalues to the real axis.

For the other quark representations with $\beta=1,4$ the joint distribution
is more subtle~\cite{AKE}. Throughout, (\ref{1}) will be assumed
for $\beta=2$,  but all results  extend to $\beta=1,2,4$ for large $N$ and/or the quenched approximation.
All units are expressed with 
$a\equiv 1$ unless noted,  which  is related to the
vacuum chiral condensate through the Banks-Casher formula~\cite{CASHER}.
Specifically $\sqrt{a}={|q^\dagger q|_0}/{\bf n}\equiv 1$,
with  ${\bf n}=N/V_4$  the density of zero modes. 

(\ref{1}) can be re-written  as an average of the complex fermion determinant

\be
Z_\beta [m_f]=\int \prod_{i=1}^Nd^2z_i\,(z_i^2+m_f^2)^{N_f}\,|{\bf \Psi}_0[z]|^2
\label{3}
\ee
using  the real many-body wave-function ${\bf \Psi}_0[z]$,
which  is the zero-mode solution to the Shrodinger equation $H_0\Psi_0=0$ with
the self-adjoint 
Hamiltonian

\be
H_0\equiv\frac 1{2m} \sum_{i=1}^{N}\left|\partial_i+{\bf a}_i\right|^2
\label{X4}
\ee
Here $\partial_i\equiv \partial/\partial z_i$ and the 
gauge potential is ${\bf a}_i\equiv \partial_iS$ with $S[z]=-{\rm ln}\Psi_0[z]$.
In (\ref{X4}) the mass parameter is $m=1/2$. 

Following \cite{HYDROUS}, we observe that the Vandermond determinant
$\Delta=\prod_{i<j}|z^2_{ij}|^{\beta}$ induces a diverging 2-body part in $H_0$.
Using a similarity transformation, we can re-absorb it in
$\Psi=\Psi_0/\sqrt{\Delta}$, and the new many-body Hamiltonian is

\be
H=\frac 1{\sqrt{\Delta}}\,H_0\,\sqrt{\Delta}
\label{X5}
\ee
We will refer to (\ref{X5}) as the quenched Hamiltonian. The phase quenched Hamiltonian follows 
a similar reasoning by rewriting (\ref{3}) as

\be
Z_\beta [m_f]=\int \prod_{i=1}^Nd^2z_i\,\left(\frac{z_i^2+m_f^2}{{\overline z}_i^2+m_f^2}\right)^{\frac{N_f}2}\,|{\bf \Psi}_f[z]|^2
\label{X3X}
\ee
Below we detail the hydrodynamical construction in the quenched approximation, and
quote the minimal changes in the phase quenched approximation for $\beta=2$.

\vskip 0.5cm
{\bf 3. Hydrodynamic.\,\,}
In the quenched approximation, we can use the collective coordinate method in~\cite{JEVICKI}  to re-write (\ref{X5}) in terms of the
density of eigenvalues as a collective variable $\rho(z) =\sum^{N}_{i=1}\delta^2(z-z_i)$.  
After some algebra and up to boundary and ultra-local terms, we obtain

\bea
H=\int d^2z \,\rho(z)\,\frac 1{2m}\left((\vec\nabla\pi)^2+({\vec{\bf A}})^2\right)\equiv \int d^2z\,{\bf h}
\label{4}
\eea
with the pair $\pi,\rho$ canonically conjugate. 
We will restrict our discussion to the 
classical limit with the pair obeying the Poisson brackets $\{\pi(z), \rho(z^\prime)\}=\delta^2(z-z^\prime)$.
Defining the even density $\rho^\chi(z)=\rho(z)+\rho(-z)$, we have

\bea
\vec{\bf A}
=\vec{A}+\frac 12 \vec\nabla\left({\beta}\rho^\chi_L(z)+(\beta-2){\rm ln}\,\sqrt{ \rho}\right)
\label{5}
\eea
Here $\rho_L$ is the logarithmic transform of $\rho$

\be
[\rho]_L\equiv \rho_L(z)=\int\, dz^\prime \,{\rm ln}|z-z^\prime|\,\rho(z^\prime)
\label{6}
\ee
and the vector potential ($\tau_\pm=1\pm \tau$)

\be
\vec{A}\equiv -\frac{N\beta}{2l^2}(\tau_-x, \tau_+ y)+\frac{\alpha}{2|z|^2}(x,y)
\label{X6X}
\ee

The equation of motion for $\rho$ yields the current conservation law 
and the Euler equation for $\vec v$. Defining $m\vec v=\vec\nabla \pi $, they
are specifically given by

\bea
\label{6X}
&&\partial_t \rho +\vec\nabla \cdot (\rho \vec{v})=0\nonumber\\
&&\partial_t \pi +\frac 12 {m {\vec v}^2} +\frac{{\vec{\bf A}}^2}{2m}\nonumber\\
&&-\frac{\beta-2}{4m\rho}{{\vec\nabla}}\cdot(\rho{\vec{\bf A}})
-\frac{\beta}{2m}{{\vec\nabla}}\cdot [\rho^\chi{\vec{\bf A}}]_L=0
\eea
The steady state flow from (\ref{6X}) corresponds to Bernoulli law
with $\partial_t\pi=C$ a fixed constant.
Note that all the relations  hold for large but finite $N$, provided that the fluid
density and velocity are  sufficiently smooth.

\vskip .5cm
{\bf 4. Hydrostatic.\,\,}
The quenched hydrostatic solution is encoded in the condition ${\bf A}(z)=0$ and $\pi=0$. 
Using the formal identity $\rho_L=(2\pi/\nabla^2)\,\rho$,  we have

\bea
\rho(z)=\frac {\kappa N}{\cal A}-\frac{\alpha}{2\beta}\delta^2(z)-\frac {\beta-2}{8\pi\beta}\nabla^2 \,{\rm ln}\,{ \rho}
\label{5X1}
\eea
where the integration constant $\kappa=1+\alpha/(2N\beta)$ is fixed by the density in leading order, 
and ${\cal A}$ is the area of the eigenvalue density.


In the phase quenched approximation for $\beta=2$, the vector potential (\ref{X6X}) is now shifted 

\be
\vec A\rightarrow \vec A+ \frac{N_f}2\vec\nabla\,{\rm ln}|z^2+m_f^2|
\ee
with the hydrodynamical equations (\ref{6X}) unchanged. The corresponding phase quenched hydrostatic 
density (\ref{5X1}) is modified 
in subleading order

\be
\rho(z)\rightarrow \rho(z) -\frac {N_f}{8\pi\beta}\nabla^2\,{\rm ln}|z^2+m_f^2|
\ee
Both the quenched and phase quenched approximations
describe an elliptic droplet at large $N$.

\vskip 0.5cm
{\bf 5. Droplet boundary.\,\,}
It is useful to recast the hydrostatic equations  in complex  form at large $N$

\bea
\label{13X}
&&\overline{z}\approx \tau z+\frac{l^2}{2N}\int_{\cal A} \frac{\rho^\chi(z^\prime)}{z-{z^\prime}}\,d^2z^{\prime }\equiv S(z)\nonumber\\
&&{z}\approx \tau \overline z+\frac{l^2}{2N}\int_{\cal A} \frac{\rho^\chi(z^\prime)}{{\overline{z}}-{{\overline z}^\prime}}\,d^2z^{\prime}
\equiv \overline{S}( \overline z)
\eea
$S(z)$ is a Schwartz function with $\overline{S}(S(z))=z$~\cite{SSZ},
i.e. $\overline{S}(z)$ is the inverse of $S(z)$. $S(z)$ is analytic outside  ${\cal A}$, with $S(z)\approx \tau z+l^2/z$
asymptotically. We note the similarity with the Blue$^\prime$s function and its use in determining
the domain boundary through conformal mapping in~\cite{BLUE}.

The support of the eigenvalues ${\cal A}$ is determined by a conic curve

\be
|z|^2+C(z^2+{\overline z}^2)+C^\prime(z+\overline{z}) +C^{\prime\prime}=0
\label{13X1}
\ee
Using the substitution $\overline{z}=S(z)$ with the asymptotic form of $S(z)$ fixes the  constants 
in (\ref{13X1}). The domain in $z=(x,y)$ is an ellipse $x^2/a_+^2+y^2/a_-^2=1$, 
with the axes

\be
\frac{a_\pm^2}{2l^2}=\frac {1\pm \tau}{1\mp \tau}
\ee
The ellipse remains un-split with
${\cal A}=\pi a_+a_-=2\pi l^2$. The 
area is preserved under $\tau$-deformation in (\ref{2}).

\vskip 0.5cm
{\bf 6. Plasmons.\,\,}
It is useful to analyze the
small deformations in the density and velocity profile by linearizing the current 
conservation law in (\ref{6X}), i.e. $\partial_t \delta \rho+{\rho_0}\nabla^2\delta \pi=0$,
which is readily solved using

\be
\delta \rho=-{\rho_0}\nabla^2 \phi\qquad\delta \pi=\partial_t \phi
\label{EX3}
\ee
Inserting  (\ref{EX3}) in the canonical action 

\be
{\bf S}=\int d^2z\,dt\,\left(\pi\partial_t \rho-{\bf h}\right)
\label{EX4}
\ee
yields  in the quadratic approximation

\be
{\bf S}\approx \int d^2z dt\,\frac{\rho_0}{2m}\left((\partial_t \vec\nabla\phi)^2-W[\phi]^2\right)
\ee
with

\be
W[\phi]=\left|\vec\nabla\left(\frac{\beta}{2}[\delta \rho]^{\chi}_L+\frac{\beta-2}{4}\frac{\delta\rho}{\rho_0}\right)\right|^2
\ee
Using again the formal identity $f_L=(2\pi /\nabla^2) f$  and defining 
the small longitudinal field $\vec\varphi\equiv\vec\nabla\phi$,  we obtain

\bea
\label{DX1}
{\bf S}\approx &&N\int d^2z dt\frac{\rho_0}{2m}\\
&&\times\left((\partial_t \vec{\varphi})^2
-\left(\frac{\pi\beta\rho_0}N\,\vec{\varphi}^{\chi}+\frac{\beta-2}{4}\nabla^2\vec{\varphi}\right)^2\right)\nonumber
\eea
after the rescaling $Nt\rightarrow t$.
The small longitudinal excitations in $\vec\varphi$ 
are gapped by the plasmon frequency $\omega_p=2\pi\beta\rho_0/N$. 
For an elliptic  droplet
of large area ${\cal A}$,  (\ref{DX1})  leads to the quadratic dispersion law

\be
\omega(k)\approx \pm \left|\omega_p-\frac{\beta-2}4 {\vec k}^2\right|
\label{DX2}
\ee
Here $|k|$ is conjugate to $|z|$. The gapped spectrum means that the droplet is uncompressible.
For $\beta=1,2$ with  quarks in the real and complex  representation the branch (\ref{DX2})
describes a plasma fluid. For $\beta=4$ with quarks in the quaternion
representation, (\ref{DX2}) shows the start of a roton-like branch a possible
indication of superfluidity. For that the higher order $k^4$ term is needed.

\vskip .5cm

{\bf 7. Odd viscosity.\,\,}There is an interesting analogy between the droplet of Dirac eigenvalues 
at finite chemical potential, and the quantum Hall effect as a fluid of 
neutralized charged electrons in the plane~\cite{STONE,WIEGMANN}. To illustrate the analogy, we first note that 
(\ref{X6X}) sources the magnetic field

\be
B(z)\equiv \vec\nabla\times{\vec{A}}^\star=\frac{N\beta}{l^2}-\pi\alpha\delta^2(z)
\label{X6X1}
\ee
with the dual 
notation ${V}_i^\star=\epsilon_{ij}{V}_j$  subsumed. Amusingly, 
(\ref{X4}) describes a Coulomb fluid in the magnetic field (\ref{X6X1}).
In large $N$ the density of eigenvalues  is uniform

\be
\rho(z)\approx \frac{N}{2\pi l^2}\approx \frac {\nu B}{2\pi}
\label{DROP}
\ee
which is the density  of  a quantum Hall droplet with filling fraction $\nu=1/\beta$. 
The plasmon frequency is the cyclotron frequency $\omega_p\equiv B/M$
with $M=N$ the analogue of the effective mass. $l$ identifies with the magnetic length.

The $k^2$-contribution in (\ref{DX2}) is  reminiscent of the odd viscosity in the 
fractional quantum Hall effect. To show this, let $\tilde\pi\equiv i\pi$ and define the collective 
velocity $m\tilde{v}=\vec\nabla\tilde\pi+{\vec{\bf A}}^\star$,
then  (\ref{4}) is a free flow-like Hamiltonian modulo ultra-local terms, 

\be
H\rightarrow \int d^2z\, \rho(z)\,\frac m{2}\,{{\tilde{v}}^+\cdot\tilde{v}}
\label{FQ2}
\ee
In our case ${\tilde{v}}^+\neq \tilde{v}$ but in the fractional quantum Hall effect they are equal,
making ${\bf A}^\star$ a real gauge-field and $\vec{\tilde v}$ a real and gauge-invariant
flow velocity for flux-riding {quasi-particles}~\cite{STONE,WIEGMANN}. Assuming
 that $\tilde\pi$ and $\rho$ are canonical and after some algebra,  the Euler equation following from
(\ref{FQ2}) yields  the momentum conservation law

\be
\partial_t(\rho\, m\tilde v_i)+\nabla_j{\bf T}_{ij}= 0
\label{7X}
\ee
with the stress tensor

\be
{\bf T}_{ij}=m\rho\,\tilde v_i\tilde  v_j+\frac{\beta-2}{4}\,\rho\, \left(\nabla_i\tilde  v_j^\star+\nabla_i^\star\tilde v_j\right)
\label{7XX}
\ee
The first contribution is the  classical free fluid part. The second contribution is 
 the odd viscosity contribution  following from the breaking of parity in 2-dimensions~\cite{AVRON}, with

\be
\frac{\eta_O}{\rho}=\frac{\beta-2}4 \rightarrow -\frac 14 , 0, \frac 12
\label{7XX1}
\ee
which is  the coefficient of the $k^2$ term in   (\ref{DX2}). In the quantum Hall fluid, 
$\eta_O$ originates from  a mixed  gauge-gravitational anomaly~\cite{MIXED}.
We note that the pair $\tilde v, {\tilde v}^\star$ are orthogonal.
This explains that the $k^2$-contribution in (\ref{DX2}) acts as
the  even (shear) viscosity but without the $i$ for dissipation. 
No vorticity is therefore expected.

\vskip 0.5cm
{\bf 8. Instanton and relaxation time.\,\,}
We identify the zero energy configuration in (\ref{4}) as an instanton solution
with imaginary velocity with $\pi\rightarrow i\pi$, i.e.

\be
{\bf h}\rightarrow |\vec\nabla\pi|^2-|{\vec {\bf A}}|^2=0
\label{XI1}
\ee
that satisfies  the analytically continued in time conservation law  ($t\rightarrow - it_E$) 

\be
-\partial_{t_E}\rho+\vec\nabla \cdot\,(\rho\vec\nabla\pi)=0
\label{IX2}
\ee
Without loss of
generality and for simplicity we choose $\tau=0$ in (\ref{2}) so that the hydrostatic droplet is circular. 
To solve (\ref{IX2}) we set $\rho(0,z)=K/\pi\gg \rho_0$,
which corresponds to all eigenvalues localized in a small disc centered around the origin to
simulate a chirally restored phase at finite $\mu$.
(\ref{IX2}) simplifies by radial symmetry

\bea
\label{IX3}
&&\partial_r \rho_L(r,t_E)=f(r,t_E)\\
&&r\partial_{t_E} f+r(\beta f-\frac{N\beta r}{2l^2})\partial_r f+f(\beta  f-\frac{N\beta r}{2l^2})=0\nonumber
\eea
We note that similar non-linear equations emerge from the diffusion of non-hermitean matrices~\cite{BGNTW}.

The solution to (\ref{IX3}) with a free boundary or large droplet size ${\cal A}$
can be obtained using the method of characteristics. Specifically

\bea
\label{IX4}
&&\frac{dt_E}{ds}=-r\nonumber\\
&&\frac{dr}{ds}=-r(\beta f-\frac{N\beta r}{2l^2})\nonumber\\
&&\frac{df}{ds}=f(\beta f-\frac{N\beta r}{2l^2})
\eea
with the conditions $t_E(s=0)=0$, $r(s=0)=r_0$ and $f(s=0)=f(r_0)$.
For $f(r,t_E=0)=Kr$ with $K\gg \rho_0$, 

\bea
\label{IX5}
&&t_E=-{\bf a}s+{\frac{l^2}{N\beta}}{{\rm ln}\left(\frac{r_0+{\bf a}-(r_0-{\bf a})e^{\frac{N\beta}{l^2}{\bf a}s}}{2{\bf a}}\right)}\nonumber\\
&&r={\bf a}\frac{r_0+{\bf a}+(r_0-{\bf a})e^{\frac{N\beta}{l^2}{\bf a}s}}{r_0+{\bf a}-(r_0-{\bf a})e^{\frac{N\beta}{l^2}{\bf a}s}}\nonumber\\
&&{\bf a}=\sqrt{2Kr_0^2l^2/N}\nonumber\\
&&f=\frac{Kr_0^2}{r}
\eea
We first use the three equations  in (\ref{IX5}) and solve for  $r_0=r_0(r,t_E)$. We then substitute the answer
in the fourth equation in (\ref{IX5}) to find explicitly $f(r, t_E)$ in general. Its large time asymptotic for
$s\rightarrow -\infty$ is $t_E\approx -{\bf a}s$ and

\be
f(r, t_E)\approx \frac{Nr}{2l^2}\left(\frac{1+\sqrt{\frac{2Kl^2}{N}}-(1-\sqrt{\frac{2Kl^2}{N}})e^{-\frac{N\beta}{l^2}t_E}}{1+\sqrt{\frac{2Kl^2}{N}}+(1-\sqrt{\frac{Kkl^2}{N}})e^{-\frac{N\beta}{l^2}t_E}}\right)^2
\label{IX6}
\ee

(\ref{IX6}) relaxes as  $e^{-2\omega_p Nt_E}$ to $f(r, \infty)=Nr/2l^2$ 
leading to the hydrostatic density $\rho_0=N/(2\pi l^2)$.
We identify $T_R\approx 1/2\omega_p$ with the relaxation time 
after rescaling  $Nt_E\rightarrow t_E$. The scalar density relaxes by emitting two  longitudinal plasmons.
In physical units 

\be
T_R\approx \frac 1{2\omega_p}\rightarrow \left(\frac{1+\frac{{\cal A}a}{2\pi}}{2\beta}\right)\,\sqrt{a}
\label{TTMU}
\ee
after re-instating $1\equiv \sqrt{a}=|q^\dagger q|_0/{\bf n}$, and adding the 1 to reproduce the $\mu=0$ result
in~\cite{HYDROUS}. A simple extension to finite temperature amounts to a re-definition of units or
$\sqrt{a}\rightarrow \sqrt{a_T}= |q^\dagger q|_T/{\bf n}_T$ as in~\cite{HYDROUS,NOTE}.

Finally,  we note that $T_R\approx 1/2\omega_p$ translates to  a diffusive time with 
$l^2\equiv{\cal A}/{2\pi}\approx 2{\beta} T_R$. The diffusion constant is ${\bf D}=2\beta$. 
An estimate of the finite droplet size corrections follow from (\ref{DX2}) using the substitution
$\omega_p\rightarrow \omega(k\approx 1/{\sqrt{\cal A}})$. The leading correction is controlled by the
odd viscosity to density ratio in (\ref{7XX1}) and is small.

\vskip 0.5cm
{\bf 9. Conclusions.\,\,}The hydrodynamical description 
captures some key aspects  of the QCD Dirac eigenvalues in the diffusive regime at finite chemical potential. It supports an
instanton  that describes  the stochastic relaxation of the Dirac eigenvalues as a fluid. The fluid is uncompressible and
exhibits dispersive plasmon waves
that can be used to estimate the time it takes for a chirally symmetric phase to relax to a chirally broken phase
in matter.  The time estimate  is non-perturbative  and gauge-independent.

\vskip 0.5cm
{\bf Acknowledgements}
The work of YL and IZ is  supported in part  by the U.S. Department of Energy under Contracts No.
DE-FG-88ER40388. The work of PW is supported by the DEC-2011/02/A/ST1/00119 grant and 
the UMO-2013/08/T/ST2/00105 ETIUDA scholarship of the (Polish) National Centre of Science.

 \vfil

\begin{thebibliography}{99}
\bibitem{NOTE}
For comparison and in pure Yang-Mills at large number of colors $N_c$,  the relaxation time for the confinement/de-confinement transition 
at finite temperature $T$, is set by the speed of sound $v_s$ in the space of holonomies, i.e.
$T_R\approx 1/{v_sT}=1/N_cT$~\cite{HYDROUS}.
\end{thebibliography}
\end{document}